\documentstyle[11pt,newpasp,twoside,epsf]{article}
\def\edcomment#1{\iffalse\marginpar{\raggedright\sl#1\/}\else\relax\fi}
\reversemarginpar

\begin{document}
\title{The ionization equilibrium of iron in \ion{H}{ii} regions}
 \author{Rodr\'\i guez, M.}
\affil{Instituto Nacional de Astrof\'\i sica, \'Optica y Electr\'onica, INAOE,
Apdo Postal 51 y 216, 72000 Puebla, Pue., Mexico; mrodri@inaoep.mx}
\author{Rubin, R. H.}
\affil{NASA Ames Research Center, Moffett Field, CA 94035--1000, USA;
rubin@cygnus.arc.nasa.gov}

\begin{abstract}
We study the ionization equilibrium of Fe using photoionization models that
incorporate improved values for the ionization and recombination cross-sections
and the charge-exchange rates for the Fe ions. The previously available
photoionization models predict concentrations of Fe$^{3+}$ which are a factor
of 3--8 higher than the values inferred from emission lines of [\ion{Fe}{iii}]
and [\ion{Fe}{iv}]. Our new models reduce these discrepancies to factors of
2--5. We discuss the possible reasons behind the remaining discrepancies and
present an updated ionization correction factor for obtaining the Fe abundance
from the Fe$^{++}$ abundance.
\end{abstract}

\section{Introduction}

Iron is expected to be in two main ionization states in \ion{H}{ii} regions:
Fe$^{++}$ and Fe$^{3+}$ (with the Fe$^+$ ion having a non negligible
concentration only for very low excitation nebulae). Therefore, the measurement
of \ion{Fe}{iii} and \ion{Fe}{iv} emission lines will allow us to determine the
Fe abundance in these nebulae. [\ion{Fe}{iii}] lines, although weak, have
already been observed in several \ion{H}{ii} regions and starburst galaxies
(e.g. Izotov \& Thuan 1999; Rodr\'\i guez 2002), but [\ion{Fe}{iv}] lines are
much weaker and difficult to observe. Therefore, the Fe abundance in \ion{H}{ii}
regions is usually obtained from [\ion{Fe}{iii}] lines and an ionization
correction factor (ICF), obtained from ionization models, to account for the
contribution of Fe$^{3+}$. For example, the relation:
\begin{equation}
\frac{\mbox{Fe}}{\mbox{O}}=\mbox{ICF}\,\frac{\mbox{Fe}^{++}}{\mbox{O}^+}=
\frac{x(\mbox{O}^+)}{x(\mbox{Fe}^{++})}\frac{\mbox{Fe}^{++}}{\mbox{O}^+},
\end{equation}
where $x(X^{n+})$ stands for the ionization fraction of the
$X^{n+}$ ion, is especially well suited for determining the Fe abundance from
optical observations of \ion{H}{ii} regions. The ionization potentials of the O
and Fe ions are similar (30.6 and 54.8 eV for Fe$^{++}$ and Fe$^{3+}$, 35.3 and
54.9 eV for O$^{+}$ and O$^{++}$). Furthermore, since both O$^{+}$ and O$^{++}$
can be measured from strong optical lines, one can get the O abundance
$\mbox{O}/\mbox{H}=\mbox{O}^{+}/\mbox{H}^{+} + \mbox{O}^{++}/\mbox{H}^{+}$, and
hence also $\mbox{Fe}/\mbox{H}$ from $\mbox{Fe}/\mbox{O}$.

However, available measurements of some weak [\ion{Fe}{iv}] lines (Rubin et al.\
1997; Rodr\'\i guez 2003) imply Fe$^{3+}$ abundances which are smaller than the
values implied by relation (1) by factors 3--8.
This discrepancy translates into
an uncertainty of up to a factor of 6 in the Fe abundances derived for a wide
range of objects, from the nearby Orion nebula to the low metallicity blue
compact galaxy SBS~0335$-$052. Thus, resolving this problem has important
implications for our understanding of both the evolution of dust in \ion{H}{ii}
regions and the chemical history of low metallicity dwarf galaxies (see the
contribution by Rodr\'\i guez \& Esteban in this volume).

In order to check whether the discrepancy is due to errors in the ICFs predicted
by models, we are studying the ionization equilibrium of Fe using
photoionization models that incorporate recently improved values for all the
atomic data relevant to the problem.

\section{Results and discussion}

We use the photoionization code NEBULA (Rubin et al.\ 1991a,b) with the
following updates: new photoionization cross sections for Fe$^{+}$, Fe$^{++}$,
O$^{0}$ and O$^{+}$ (Nahar \& Pradhan 1994; Nahar 1996a; Kjeldsen et al.\ 2002;
Verner et al.\ 1996), new recombination coefficients for Fe$^{++}$, Fe$^{3+}$,
O$^{+}$ and O$^{++}$ (Nahar 1996b, 1997, 1999), all the charge-exchange
reactions involving these ions (Kingdon \& Ferland 1996; and the ORNL/UGA Charge
Tranfer Database for Astrophysics ---
http://www-cfade.phy.ornl.gov/astro/ps/data); and the NLTE model stellar
atmospheres of Pauldrach et al.\ (2001) with solar metallicity.

The ionization cross sections are not smoothly varying; they show several
resonances at different energies. Since the positions of these resonances have
uncertainties of a few percent, we smoothed (or used available smooth fits) both
the ionization cross sections and the model atmospheres by convolving with
Gaussians of widths 3\% and 1.5\% (in energy), respectively.

Figure~1 shows the ICFs obtained from different models with
$T_{\rm eff}=35000$--50000~K, total nucleon density $N=100$--10000~cm$^{-3}$,
and Orion metallicity Z, Z/10, and Z/30. The results of previous ionization
models are also shown for comparison. The model results are compared with the
results obtained by Rodr\'\i guez (2003) for several objects with available
measurements of [\ion{Fe}{iv}] lines. The new model results, although still
higher than the measured values, are much closer to them. The remaining
discrepancies could be due to: (1) the need for further improvements in the
photoionization models, (2) errors in the atomic data (in particular the
collision strengths) used to derive the Fe$^{++}$ and Fe$^{3+}$ abundances. The
effect on the calculated values of $x(\mbox{O}^+)/x(\mbox{Fe}^{++})$ of either
decreasing the Fe$^{++}$ abundance or increasing the Fe$^{3+}$ abundance by a
factor of 2 is also shown in Figure~1. This change would make the results for
IC4846, 30 Doradus (its upper limit), and M42 compatible with the model
results. The possibility of such a change is provided by the recent
calculations of collision strengths for Fe$^{++}$ by McLaughlin et al.\ (2002).
However, they only calculate the collision strengths for transitions between
terms. Hence, the new collision strengths cannot be used either to derive
abundances or to check their reliability by comparing their predictions with
the observed relative intensities of various [\ion{Fe}{iii}] lines.

\begin{figure}[ht]
\plotone{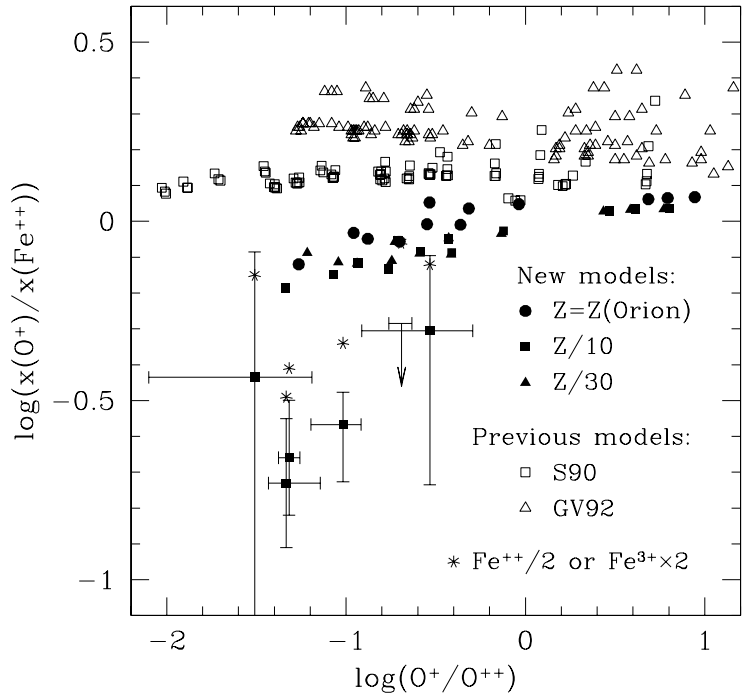}
\caption{ICFs obtained from our new models compared with the previous model
results of Stasi\'nska (1990, S90) and Gruenwald \& Viegas (1992, GV92). The
squares with error bars show the results obtained in Rodr\'\i guez (2003) for
objects with available measurements of [\ion{Fe}{iv}] lines: from left to right,
the planetary nebula IC4846, SMC~N88A (one position in the object), the low
metallicity blue compact galaxy SBS~0335$-$052, SMC~N88A (a second position), 30
Doradus (an upper limit), and M42 (the Orion nebula). The asterisks show the
effect on the calculated values of either decreasing the Fe$^{++}$ abundance or
increasing the Fe$^{3+}$ abundance by a factor of 2.}
\end{figure}

Even if the factor of 2 decrease in the Fe$^{++}$ abundance proved to be real,
it would not explain the discrepancies shown by the objects with lower
metallicities, SMC~N88A and SBS~0335$-$052. However, the models with lower
metallicity show slightly lower values for the ICF; thus further refinements
(such as the use of low metallicity model atmospheres) might improve the fit. On
the other hand, since different [\ion{Fe}{iv}] lines were used to derive the
Fe$^{3+}$ abundance for  the five objects, changes in the relevant collision
strengths might explain all the discrepancies.

In any case, the Fe abundances implied by [\ion{Fe}{iii}] emission lines should
be obtained through the relation implied by a fit through the new model results
in Figure~1:
\begin{equation}
\frac{\mbox{Fe}}{\mbox{O}}=\Bigg(\frac{\mbox{O}^+}{\mbox{O}^{++}}\Bigg)^{0.09}
	\,\frac{\mbox{Fe}^{++}}{\mbox{O}^+}.
\end{equation}

\acknowledgments{Project partially supported by the Mexican CONACYT project
J37680-E. RR acknowledges support from NASA Long-Term Space Astrophysics
program.}


\begin{references}
\reference Gruenwald, R.B., \& Viegas, S.M. 1992, \apjs, 78, 153 (GV92)
\reference Izotov, Y.I., \& Thuan, T.X. 1999, \apj, 511, 639
\reference Kingdon, J.B., \& Ferland, G.J. 1996, \apjs, 106, 205
\reference Kjeldsen, H., Kristensen, B., Folkman, F., \& Andersen, T. 2002, J.
	Phys. B, 35, 3655
\reference McLaughlin, B.M., Scott, M.P., Sunderland, A.G., Noble, C.J., Burke,
	V.M, \& Burke, P.G. 2002, J. Phys. B, 35, 2755
\reference Nahar, S.N. 1996a, \pra, 53, 1545
\reference Nahar, S.N. 1996b, \pra, 53, 2417
\reference Nahar, S.N. 1997, \pra, 55, 1980
\reference Nahar, S.N. 1999, \apjs, 120, 131
\reference Nahar, S.N., \& Pradhan, A.K. 1994, J. Phys. B, 27, 429
\reference Pauldrach, A.W.A., Hoffman, T.L., \& Lennon, M. 2001, \aap, 375, 161
\reference Rodr\'\i guez, M. 2002, \aap, 389, 556
\reference Rodr\'\i guez, M. 2003, \apj, 590, 296
\reference Rubin, R.H., Simpson, J.P., Haas, M.R., \& Erickson, E.F. 1991a,
	\apj, 374, 564
\reference Rubin, R.H., Simpson, J.P., Haas, M.R., \& Erickson, E.F. 1991b,
	\pasp, 103, 834
\reference Rubin, R.H. Dufour, R.J., Ferland, G.J., Martin, P.G., O'Dell, C.R.,
	Baldwin, J.A., Hester, J.J., Walter, D.K., \& Wen, Z. 1997, \apj, 474,
	L131
\reference Stasi\'nska, G. 1990, \aaps, 83, 501 (S90)
\reference Verner, D.A., Ferland, G.J., Korista, K.T., \& Yakolev, D.G. 1996,
	\apj, 465 487
\end{references}
\end{document}